\documentclass[12pt,a4paper]{article}
\usepackage{epsfig}
\begin{document}

\title{  
{\Large \bf Higher Order Effects in Non Linear Evolution from a Veto in Rapidities}  }  
\author{  
{\large G. Chachamis \thanks{e-mail: chachami@mail.desy.de}~~$\mathbf{{}^{a}}$, M. ~Lublinsky\thanks{e-mail: lublinm@mail.desy.de}~~$\mathbf{{}^{b}}$, 
A. Sabio~Vera\thanks{e-mail: sabio@mail.desy.de}~~$\mathbf{{}^{a}}$  
}\\[4.5ex]  
{\it ${}^{a}$ II. Institut f{\"u}r Theoretische Physik, Universit{\"a}t 
Hamburg,}\\{\it Luruper Chaussee 149, 22761~Hamburg, Germany}\\\\[2.5ex]  
{\it ${}^{b}$ DESY Theory Group, 22603, Hamburg, Germany}\\[4.5ex]  
}  
  
\maketitle

\vspace{-14cm}
\begin{flushright}  
DESY 04-153
\end{flushright}  

\vspace{14cm}
\begin{abstract}  
Higher order corrections to the Balitsky--Kovchegov equation 
have been estimated by introducing a rapidity veto which forbids subsequent 
emissions to be very close in rapidity and is known to mimic higher order 
corrections to the linear BFKL equation. The rapidity veto constraint has  
been first introduced using analytical arguments obtaining a power growth 
with energy, $Q_s ({\rm Y}) \sim e^{\lambda {\rm Y}}$, of 
the saturation scale of $\lambda \sim 0.45$. Then a numerical analysis for 
the non--linear 
Balitsky--Kovchegov equation has been carried out for phenomenological 
rapidities: when a veto of about two units of rapidity is introduced 
for a fixed value of the coupling constant of $\alpha_s = 
0.2$ the saturation scale $\lambda$ decreases 
from $\sim 0.6$ to $\sim 0.3$, and when running coupling effects are 
taken into account it decreases from $\sim 0.4$ to $\sim 0.3$.
\end{abstract}  

\newpage

\section{Introduction}  

The high energy behavior of a parton system can be associated to 
the Balitsky--Fadin--Kuraev-Lipatov (BFKL) dynamics \cite{FKL}. 
At leading order (LO) the BFKL equation resums contributions of the form 
$(\alpha_s\,{\rm Y})^n$, with ${\rm Y} \sim \ln s$ being a rapidity variable.  
Linear evolution gives rise to a Pomeron--like behavior of the  
scattering amplitudes 
with an intercept $\omega^{\rm BFKL} \simeq 0.5$. This power growth of the
amplitude with energy violates $s$-channel unitarity at rapidities 
of the order of ${\rm Y} \,\sim \,1/\alpha_s \,\ln \,1/\alpha_s$ \cite{MU97}. 

A theoretical possibility for the high energy growth of the amplitudes to 
be modified in a way consistent with unitarity is the idea of parton 
density saturation \cite{GLR}, which accounts for the possibility of 
parton annihilation, an essentially nonlinear effect.
 Present theoretical understanding
views a system of saturated partons as a new state of matter called 
Color Glass Condensate (CGC) (see e.g. Ref.~\cite{Mc} and references therein).
Saturation signals have been intensively searched 
experimentally mostly at HERA and more recently at RHIC. 
Although there are hints of parton density saturation in the present data, 
a clean and ultimate signal has not been found so far.

The fundamental quantity characterizing the transition to the saturation 
regime is the so--called ``saturation
scale'', $Q_s({\rm Y})$. The determination of the rate of growth with rapidity 
of this saturation scale could be of a large importance for, e.g., structure
function extrapolations from HERA to LHC kinematics. In the context of the 
well--known saturation model of Golec-Biernat and Wusthoff (GBW) 
\cite{Golec-Biernat:1998js}, the saturation scale grows exponentially as 
$Q_s ({\rm Y}) \sim \exp(\lambda\,{\rm Y}/2)$ with $\lambda \simeq 0.29$. 

Within the LO approximation a theoretical tool with solid grounds in 
perturbative QCD suitable to study saturation phenomena at high energies is 
the Balitsky--Kovchegov nonlinear evolution equation (BKe)
\footnote{Eq.~(\ref{EQ}) was originally  
proposed by Gribov, Levin and Ryskin \cite{GLR} in momentum space  
and derived in the double logarithmic approximation by  
Mueller and Qiu \cite{MUQI}. In the leading $\ln 1/x$  
approximation it was obtained by Balitsky using a Wilson Loop  
Operator Expansion \cite{BA}. In the form presented in Eq.~(\ref{EQ}) it  
was obtained by Kovchegov \cite{KO} using the color dipole approach  
\cite{MU94} to high energy scattering in QCD. This equation was  
also obtained by summation of BFKL Pomeron fan diagrams by  
Braun \cite{Braun} and, more recently, 
Bartels, Lipatov and Vacca \cite{BLV}.   
In the framework of the Color Glass Condensate it was obtained  
by Iancu, Leonidov  and McLerran \cite{ILM}}. This equation reads  
\begin{eqnarray}  
 \frac{dN({\mathbf{x_{01}}},{\rm Y})}{d {\rm Y}} = \bar{\alpha}_s 
 \int_{\rho} \frac{d^2 {\mathbf{x_{2}}}}{2\,\pi}   
\frac{{\mathbf{x^2_{01}}}}{{\mathbf{x^2_{02}}}  
{\mathbf{x^2_{12}}}} 
\left[2 N({\mathbf{x_{02}}},{\rm Y}) - N({\mathbf{x_{01}}},{\rm Y}) - 
N({\mathbf{x_{02}}}, {\rm Y}) N({\mathbf{x_{12}}},{\rm Y})\right], 
\label{EQ}
\end{eqnarray}  
with $\bar{\alpha}_s \equiv \alpha_s N_c /\pi$. In 
the color dipole approach to high energy scattering 
the function $N(r_{\perp},{\rm Y},b)$ 
stands for the  imaginary part of the amplitude for a dipole of size  
$r_{\perp}$ elastically scattered at an impact parameter $b$.  
In this paper the impact parameter dependence
of the amplitude will be neglected, considering, in this way, a target of 
infinite size. $\rho$ is an ultraviolet cutoff needed to regularise the  
integral which does not appear in physical quantities. 

The physical 
content of Eq.~(\ref{EQ}) is that of a dipole of size  
$\mathbf{x_{10}}$ which decays into two other dipoles of size  
$\mathbf{x_{12}}$ and $\mathbf{x_{02}}$  with a decay probability  
given by the wave function  $| \Psi|^2  
\,=\,\frac{\mathbf{x^2_{01}}}{\mathbf{x^2_{02}}\,\mathbf{x^2_{12}}}$.  
 These two dipoles can then interact with the target and, with a certain 
probability, they can do so simultaneously, a possibility accounted for by 
the non--linear term in Eq.~(\ref{EQ}). 
The linear part in Eq.~(\ref{EQ}) corresponds to the LO BFKL equation 
describing the evolution with energy of the multiplicity of the fixed  
size color dipoles. 
The BKe has been studied both analytically \cite{LT,KW,IIM,PM}
and numerically \cite{Braun,GLLM,Braun2,GMS,LGLM,Weigert,BKEb,KS,AW}. 
Phenomenologically the BKe provides a good description of DIS 
data from HERA~\cite{GLLM,GLLMN,L,Iancu}. It is worth pointing out that 
the linear part of the BKe is obtained in the leading soft gluon emission 
approximation keeping the strong coupling fixed and that the large $N_c$ 
limit is used in order to write the nonlinear term as a product of 
two functions $N$. This limit is at the basis of the color dipole picture 
and, to a large extent, it corresponds to a mean field theory without dipole 
correlations. The equation also neglects target correlations, an assumption 
which might be valid for asymptotically heavy nuclei but not for protons or a 
realistic nucleus target.

It would be very interesting to go beyond the original BKe and relax some 
of the underlying assumptions outlined above. At present there is a large 
activity in this direction. Regarding the contribution of the $N_c$ corrections
they can be estimated to be   up to 15\% \cite{Weigert}.
In this publication we would like to 
focus on the higher order $\alpha_s$ corrections which are relevant, in 
particular, for phenomenological applications. 

In principle, unitarity corrections based on LO estimates 
are expected to be important at rapidities of the order
${\rm Y}\,\sim \,1/\alpha_s \,\ln \,1/\alpha_s$,  
parametrically earlier than the   
next--to--leading (NLO) corrections which set in 
at ${\rm Y}\,\sim \,1/\alpha_s^2$. 
It is also known that the NLO corrections to the linear BFKL equation 
significantly decrease the Pomeron intercept thus postponing the arrival 
of unitarity corrections to higher rapidities. 

A complete nonlinear equation at NLO has not been derived yet. 
In the conventional approach based on $s$-channel unitarity,
the forward BFKL kernel is known at NLO~\cite{Fadin:1998py,Ciafaloni:1998gs}. 
A nonlinear evolution needs the knowledge of the non-forward 
kernel~\cite{Fadin:2000nq} together with the NLO impact 
factor~\cite{Fadin:2002tu,Fadin:1999de,Bartels:2004bi}, both calculations 
being currently under investigation. However,
a NLO study of the triple 
Pomeron vertex entering the BKe has not been initiated yet. So far, the only 
exact result which has been reported is due to      
 Balitsky and  Belitsky~\cite{BB} who have been able to compute a single
NLO contribution with maximal nonlinearity, the $N^3$ term.  

There have been some attempts to get insight about saturation at NLO 
using approximate methods. Triantafyllopoulos~\cite{DT}  has considered 
the renormalisation group improved NLO BFKL equation with the  
presence of a saturation boundary. His results show a decrease in the 
saturation scale 
growth as a function of rapidity towards the value $\lambda\,\simeq \,0.3$ 
observed experimentally.  A similar type of study
based on the NLO BFKL has been recently reported in~\cite{Khoze:2004hx}. 

In this work we propose a new approach for 
the study of saturation effects including NLO corrections. We will introduce 
a constraint in the rapidity of the emitted gluons in the BKe, a so--called 
``rapidity veto''~\cite{Schmidt:1999mz,Forshaw:1999xm} which, for the linear part of the equation, is known to reproduce the bulk of the NLO corrections. 
In the next section we revise how to introduce a veto in rapidity in the 
linear BFKL equation and apply this constraint to obtain an estimate of 
the saturation scale as a function of the veto. In Section~\ref{nnBKe} we 
apply the method of rapidity veto to the BKe and study its influence on the 
energy growth of the saturation scale. In the last Section of this work 
we present our summary.

\section{The rapidity veto in BFKL and the saturation line}
\label{RapidityVeto}

\begin{figure}
\centerline{\epsfig{file=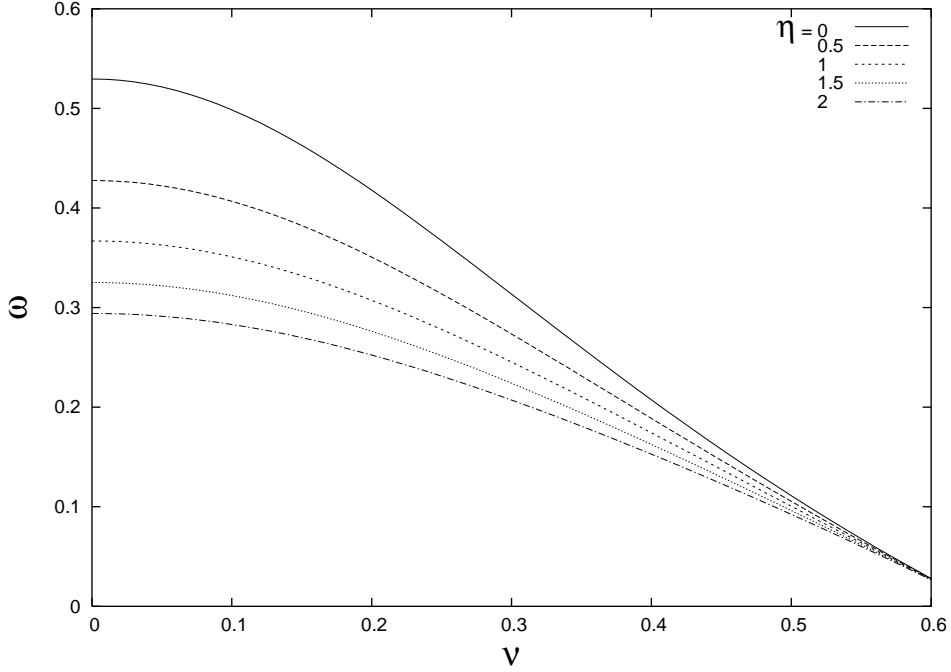,width=9cm,angle=-90}}
\caption{Dependence of the LO eigenvalue of the kernel on the veto upon 
$\nu$}
\label{VetoLO}
\end{figure}

In the following the introduction of a rapidity veto as in 
Ref.~\cite{Schmidt:1999mz,Forshaw:1999xm} will be shown. To impose 
the constraint that subsequent gluon emissions are separated by some 
minimum interval in rapidity, $\eta$, can be done writing the LO BFKL equation 
as an integral equation in rapidity, i.e.
\begin{eqnarray}
f({\rm Y},\gamma) &=& \int d {\rm Y}' \, \theta({\rm Y}-{\rm Y}' - \eta) 
\,\bar{\alpha}_s \, \chi(\gamma) \, f({\rm Y}',\gamma),
\end{eqnarray}
where $\gamma$ corresponds to a Mellin transform in transverse momentum 
space and $\chi (\gamma) = 2 \Psi(1) - \Psi(\gamma) - \Psi(1-\gamma)$ 
is the eigenvalue of the LO kernel. To go to the representation in 
the $\omega$ plane we use the transformation
\begin{eqnarray}
f_{\omega}(\gamma) &=& \int d{\rm Y} \, e^{-\omega {\rm Y}} \, 
f ({\rm Y},\gamma). 
\end{eqnarray}
We can now introduce this in the original equation to obtain
\begin{eqnarray}
f_{\omega}(\gamma) &=& \bar{\alpha}_s \, \chi(\gamma) \, 
\int d{\rm Y}' \, f({\rm Y}',\gamma) \, \int_{{\rm Y}'+\eta} d{\rm Y}\,
e^{-\omega {\rm Y}},
\end{eqnarray}
therefore the effect of imposing the veto on the LO BFKL equation leads 
to an eigenvalue which is determined by the solution to 
\begin{eqnarray}
\omega &=& \bar{\alpha}_s \chi(\gamma) \, e^{- \eta \omega}.
\label{vetito}
\end{eqnarray}
It is worth noting that the solution to this equation respects the 
structure of a maximum at $\gamma = \frac{1}{2}+ {\rm i} \, \nu$ for 
$\nu \simeq 0$ so that this region dominates at high energies. This is 
highlighted in Fig.~\ref{VetoLO} where the maxima are shown revealing how 
the original value of the Pomeron intercept decreases from about 0.5 to 
about 0.3 for a value of the veto of two units of rapidity. This is in 
agreement with other predictions from studies of the NLO gluon Green's 
function~\cite{NLLpapers}.

\begin{figure}
\centerline{\epsfig{file=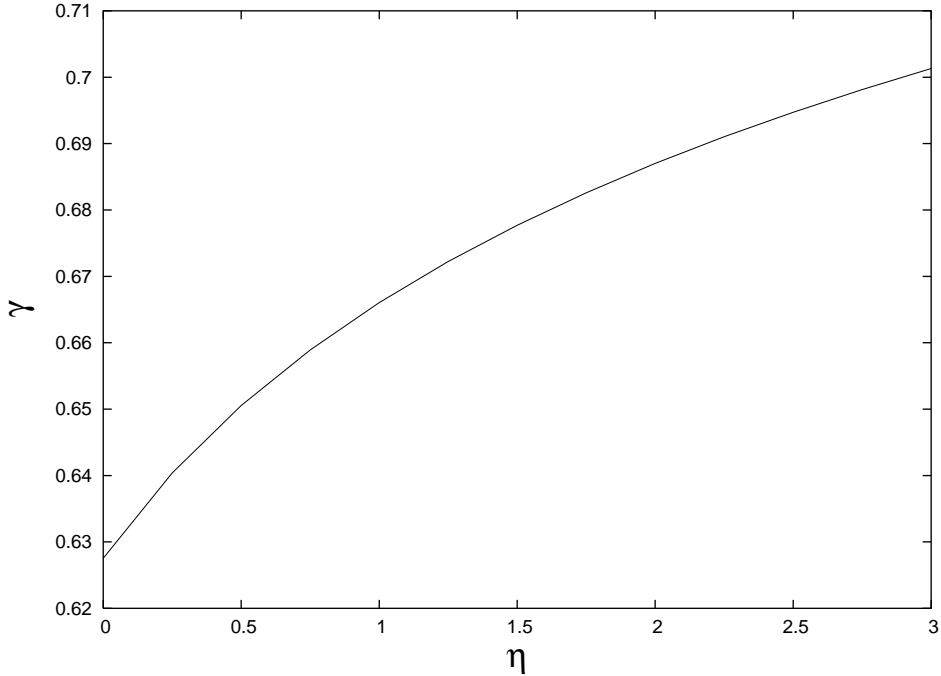,width=9cm,angle=-90}}
\caption{The solution to Eq.~(\ref{critical_solution}) as a function of the veto in rapidity, $\eta$.} 
\label{Gamma_veto}
\end{figure}

Now we calculate the influence of this veto in rapidity on the saturation 
scale. In the case of forward scattering the amplitude for a dipole of 
size $1/Q$ on a dipole of size $1/Q_0$ can be written as 
\begin{eqnarray}
\mathcal{N} \left(Q,Q_0,{\rm Y}\right) &=& \int d \gamma 
\int d \omega \, \mathcal{N}_0 \left(\gamma \right)
\exp{\left(-\gamma L + \omega {\rm Y}\right)}\frac{1}{\omega- {\bar \alpha}_s 
\chi \left(\gamma\right)},
\end{eqnarray}
with $L \equiv \ln{Q^2/Q_0^2}$. The veto in rapidity is easily introduced via 
a modified kernel as in Eq.~(\ref{vetito}):
\begin{eqnarray}
\mathcal{N} \left(Q,Q_0,{\rm Y},\eta\right) &=& \int d \gamma
\int d \omega \, \mathcal{N}_0 \left(\gamma \right)
\exp{\left(-\gamma L + \omega {\rm Y}\right)}\frac{1}{\omega- {\bar \alpha}_s 
\chi \left(\gamma\right) e^{- \eta \, \omega}}.
\end{eqnarray}
The saturation line, $L_s \equiv \ln{Q_s^2({\rm Y}) /Q_0^2}$ with $Q_s^2 ({\rm Y}=0) \equiv Q_0^2$, 
can be defined as that with a stationary exponent:
\begin{eqnarray}
- \gamma L_s + \omega(\gamma, \eta) {\rm Y} = 0, 
\end{eqnarray}
where the introduction of the veto enforces
\begin{eqnarray}
\omega(\gamma, \eta) &=& {\bar \alpha}_s \chi(\gamma) \exp{(- \eta \, \omega (\gamma,\eta))}.
\end{eqnarray}
At high energies the dominant region is that in the intersection with 
the saddle point ${\bar \gamma}$
\begin{eqnarray}
- L_s + \left.\frac{\omega(\gamma,\eta)}{d \gamma}\right|_{\gamma = \bar{\gamma}} {\rm Y} &=& 0.  
\end{eqnarray}
The solution to this system of equations provides an implicit equation for 
${\bar \gamma}$:
\begin{eqnarray}
\frac{\chi'({\bar \gamma})}{\chi({\bar \gamma})}{\bar \gamma} -1 
&=& {\bar \alpha}_s \, \eta \, {\chi({\bar \gamma})} 
\exp{\left(1- \frac{\chi'({\bar \gamma})}{\chi({\bar \gamma})}{\bar \gamma}
\right)}.
\label{critical_solution}
\end{eqnarray}
Consequently, when the rapidity veto is imposed it develops a dependence 
on the ${\bar \alpha}_s \eta$ product, 
${\bar \gamma} = {\bar \gamma} \left({\bar \alpha}_s \eta\right)$, and the 
saturation line reads now
\begin{eqnarray}
L_s &=& {\bar \alpha}_s \frac{\chi({\bar \gamma})}{\bar \gamma}\,{\rm Y}\, 
\exp{\left(1- \frac{\chi'({\bar \gamma})}{\chi({\bar \gamma})}{\bar \gamma} 
\right) } \equiv \lambda( \bar{\alpha}_s, \eta) \, {\rm Y}.
\label{Ls}
\end{eqnarray} 
For a value of $\alpha_s = 0.2$ in Fig.~\ref{Gamma_veto} we have 
plotted the dependence of ${\bar \gamma}$ on the rapidity veto $\eta$. The 
effect of the constraint in rapidity 
is to increase the value of this critical 
$\bar{\gamma}$ from the well known $\bar{\gamma} \simeq 0.63$ (a value which 
is obtained for $\eta \rightarrow 0$ and/or in the limit of $\bar{\alpha}_s \rightarrow 0$) to about 0.69 for $\eta \sim 2.5$. 
This is in agreement with the recent results for this 
quantity of Ref.~\cite{Khoze:2004hx} where a resummed NLL BFKL equation 
was under study.

\begin{figure}
\centerline{\epsfig{file=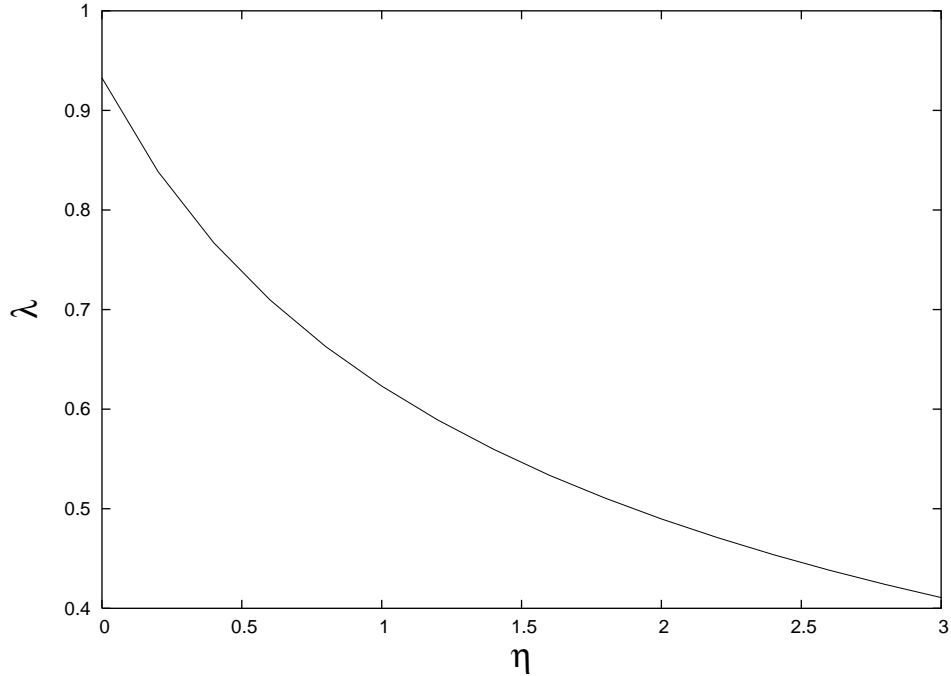,width=9cm,angle=-90}}
\caption{The saturation power (Eq.~(\ref{Ls})) as a function of the veto in rapidity, $\eta$.} 
\label{Lambda_veto}
\end{figure}

The evolution in energy is determined by the  
$d L_s / d {\rm Y} = \lambda (\bar{\alpha}_s, \eta)$ derivative. 
This $\lambda$ term is calculated in Fig.~\ref{Lambda_veto} where it can be seen how the 
effect of the rapidity constraint is to delay the onset of the saturation line by 
means of decreasing $\lambda$ at larger values of the veto. For 
zero veto it corresponds to the usual value of this linear coefficient of 
$\lambda \sim 0.93$, which can be calculated from Eq.~(\ref{critical_solution}) setting $\eta = 0$, $\alpha_s = 0.2$ and then introducing the result for 
${\bar \gamma}$ in Eq.~(\ref{Ls}) reading 
$\lambda = {\bar \alpha}_s \chi({\bar \gamma})/ {\bar \gamma}$. 
At a rapidity constraint of $\eta = 2.5$ we obtain  
$\lambda \simeq 0.45$, in agreement with Ref.~\cite{Khoze:2004hx} 
and larger than that calculated in Ref.~\cite{DT}. 

These results have been produced in the case of linear evolution imposing 
a constraint which reproduces higher order corrections and estimating the 
position of the saturation line. In the following section we introduce the 
veto in rapidities directly in the non--linear BKe and study the consequences 
of this constraint on the evolution for phenomenological rapidities. We will see that the power growth of the saturation scale is slower, even for zero veto, 
mainly due to preasymptotic effects. We will then show how the effect of 
the veto in rapidity is very similar to that found in this section: the 
effective power decreases as the veto is larger.

\section{The rapidity veto in the BKe}
\label{nnBKe}

\subsection{The fixed coupling case}

To proceed with the numerical analysis it is convenient to write the BKe 
for fixed coupling in the integral form
\begin{eqnarray}    
 N({\mathbf{x_{01}}},{\rm Y}) &=& N({\mathbf{x_{01}}},{\rm Y}_0)\,+\,
\, \bar{\alpha}_s \, \int_{{\rm Y}_0}^{\rm Y}d\,{\rm Y}'\,
 \int_{\rho} \, \frac{d^2 {\mathbf{x_{2}}}}{2\,\pi}   
\frac{{\mathbf{x^2_{01}}}}{{\mathbf{x^2_{02}}}\,  
{\mathbf{x^2_{12}}}} \,\,\times \nonumber\\  
& &\left[\,2\,  N({\mathbf{x_{02}}},{\rm Y}')\,- 
\,N({\mathbf{x_{01}}},{\rm Y}')\,-\, 
N({\mathbf{x_{02}}},{\rm Y}')\,  N({\mathbf{x_{12}}},{\rm Y}')\right], 
\label{BKEintegral}
\end{eqnarray} 
with the initial condition being defined at the rapidity ${\rm Y}_0$. As 
it was said above the rapidity veto prevents two emissions from being 
emitted close to each other in rapidity space. In this work we 
impose this veto in both the linear and non--linear parts of the BKe. In 
this way we maintain the locality of the recombination process, corresponding 
to the quadratic term. Hence, the new equation simply reads
\begin{eqnarray}    
 N({\mathbf{x_{01}}},{\rm Y}) &=& N({\mathbf{x_{01}}},{\rm Y}_0)\,+\,
\, \bar{\alpha}_s \, \int_{{\rm Y}_0+\eta}^{\rm Y-\eta}d\,{\rm Y}'\,
 \int_{\rho} \, \frac{d^2 {\mathbf{x_{2}}}}{2\,\pi}   
\frac{{\mathbf{x^2_{01}}}}{{\mathbf{x^2_{02}}}\,  
{\mathbf{x^2_{12}}}} \,\,\times \nonumber \\  
& &\left[\,2\,  N({\mathbf{x_{02}}},{\rm Y}')\,- 
\,N({\mathbf{x_{01}}},{\rm Y}')\,-\, 
N({\mathbf{x_{02}}},{\rm Y}')\,  N({\mathbf{x_{12}}},{\rm Y}')\right]. 
\label{BKEintegralveto}
\end{eqnarray} 

In the numerical implementation we work with the differential form of this 
equation with veto, i.e.,
\begin{eqnarray}    
 \frac{d N({\mathbf{x_{01}}},{\rm Y})}{d {\rm Y}} &=& 
\bar{\alpha}_s \, \int_{\rho} \, \frac{d^2 {\mathbf{x_{2}}}}{2\,\pi}    
\frac{{\mathbf{x^2_{01}}}}{{\mathbf{x^2_{02}}}\, {\mathbf{x^2_{12}}}} \times
 \nonumber\\
&&\hspace{-1.8cm}\left[\,2\,  N({\mathbf{x_{02}}},{\rm Y}-\eta)\,- 
\,N({\mathbf{x_{01}}},{\rm Y}-\eta)\,-\, 
N({\mathbf{x_{02}}},{\rm Y}-\eta)\,  N({\mathbf{x_{12}}},{\rm Y}-\eta)\right], 
\label{BKEdifveto}
\end{eqnarray} 
which highlights the non--locality in rapidity after the constraint has been 
imposed. 
Expanding Eq.~(\ref{BKEdifveto}) in $\eta$ (we assume $\eta\,\ll\,\,\rm Y$)
one can easily verify
that the corrections introduced via veto are proportional to $\alpha_s^2$ 
and thus of the NLO.

In the context of the BKe the imposition of the veto has the 
consequence of a small fluctuation above $N = 1$ when the function 
approaches the unitarity bound. These small fluctuations do not grow with 
rapidity. Technically this small violation of unitarity has its origin 
in the fact that the evolution is not stopped at $N({\rm Y}) = 1$ since the 
derivative is computed at a retarded rapidity ${\rm Y} - \eta$, see 
Eq.~(\ref{BKEdifveto}). This small instability is not surprising 
since our approach is only an estimate of the NLO corrections.
To fully preserve unitarity 
possibly we would have to introduce a correlation between coordinates and 
rapidity. 
Connecting with this point 
it is worth noticing that a generalization of the BKe proposed in 
Ref. \cite{LL} and amounting to having extra $1-N({\mathbf{x_{01}}},Y)$ 
factor in front of the evolution kernel would respect unitarity 
even in the presence of a rapidity veto. In  order to study the behavior 
of the saturation scale, 
our analysis will be centered around the transition region $N \simeq 0.5$,
which is a region not affected by the above issue.

Eq.~(\ref{BKEdifveto}) will be solved numerically for $x\le x_0= 0.01$,  
which corresponds to rapidities above ${\rm Y}_0 \simeq 4.65$.
With this goal in mind we need to specify
the initial conditions to the non--linear equation which, in principle, 
should be fitted to experimental data. Motivated by the phenomenological 
accuracy of the results in Ref.~\cite{GLLM}, the same initial conditions 
as those in that reference are used in the present work. These conditions were 
fitted to low $x \, F_2$ data for the BKe with no veto and read
\begin{eqnarray}
N(r,{\rm Y}_0) &=& 1\,-\,\exp{\left(-\alpha_s \,C_F\,r^2\,x\, G^{\rm CTEQ}/(\pi R^2)\right)}.
\label{ini}
\end{eqnarray}
Here  $\alpha_s$ is taken to be LO running at the scale $4/r^2$, and   
$x \, G^{\rm CTEQ}$ is the LO CTEQ6  gluon distribution also computed at the scale $4/r^2$. The initial condition in~(\ref{ini}) is smoothly extrapolated to $N=1$ 
at very large distances using the method proposed in Ref.~\cite{L} and
implemented in Ref.~\cite{GLLM}. The parameter $R$ stands for the
effective proton size, $R^2=3.1\,{\rm GeV}^{-2}$, an output of the fit 
performed
in Ref.~\cite{GLLM}. 
For the numerical implementation of the veto it is necessary to 
generate the initial conditions in a band 
of width $\eta$ between ${\rm Y}_0$ and ${\rm Y}_0-\eta$. 
The reason for this becomes clear if we think the evolution
in rapidity ${\rm Y}$ as a process in which  $N(r,{\rm Y})$ 
 is the input for the next step which will give us
$N(r,{\rm Y}+\delta {\rm Y})$. In the non zero veto case 
in order to compute $N(r,{\rm Y}+\delta {\rm Y})$ one would
need to consider $N(r,{\rm Y}-\eta +\delta {\rm Y})$. Hence, since
we start at ${\rm Y}_0$ we need a band of initial conditions
that will span the space between  ${\rm Y}_0- \eta$ and ${\rm Y}_0$.
As there is no evolution in energy along that band we assume the
initial conditions to be independent of ${\rm Y}$ and equal 
to $N(r,{\rm Y}_0)$ on the band.

\begin{figure}[htbp]
 \epsfig{file=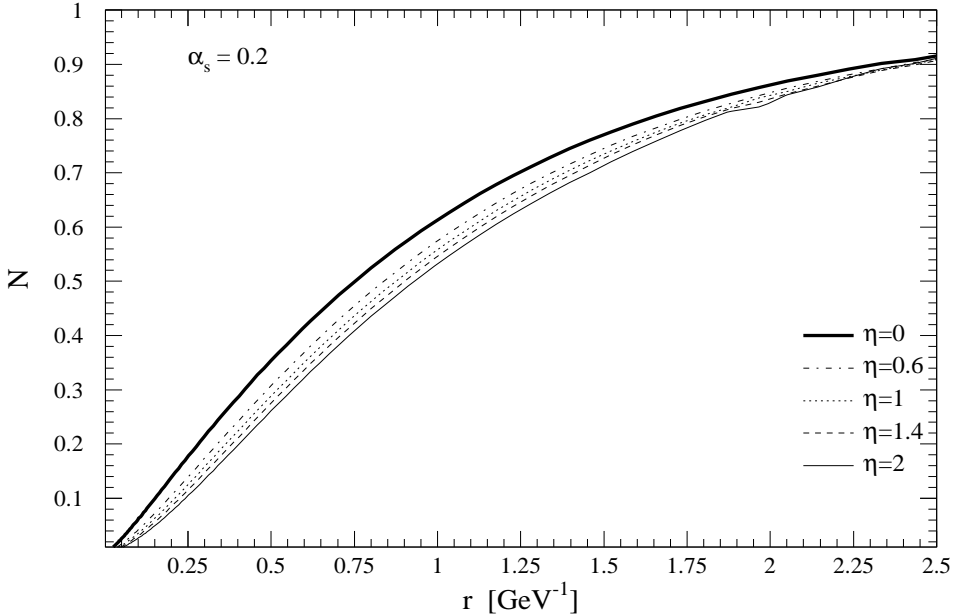,width=140mm, height=100mm}
\vspace{-0.5cm}
 \caption{Solution to BKe for different values of the veto 
       as a function of $r$, for ${\rm Y}=10$ and $\alpha_s=0.2$.}
\label{4}
\end{figure}

\begin{figure}[htbp]
 \epsfig{file=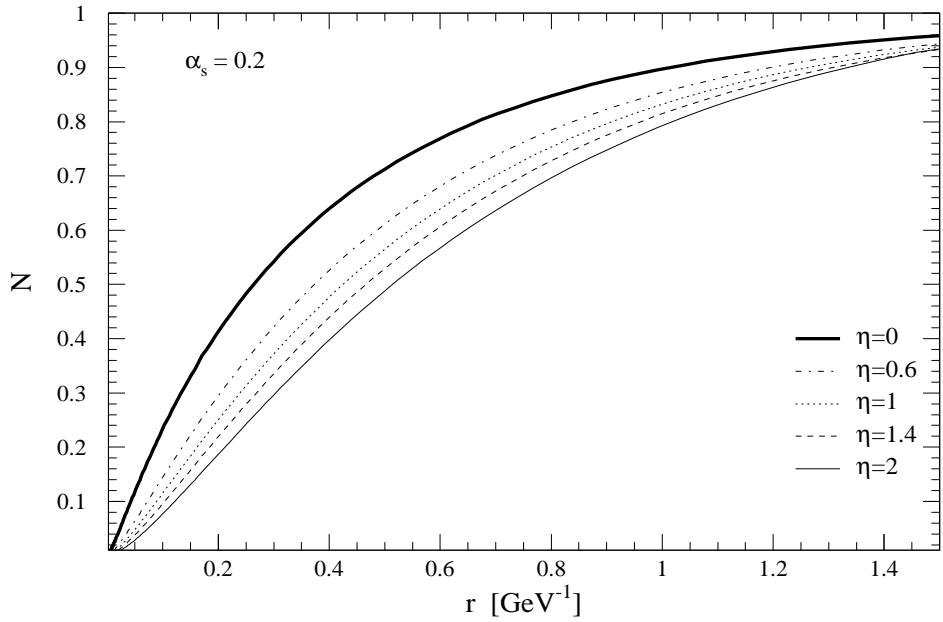,width=140mm, height=100mm}
\vspace{-0.5cm}
 \caption{Solution to BKe for different values of the veto 
as a function of $r$, for ${\rm Y}=14$ and $\alpha_s=0.2$ }
\label{5}
\end{figure}

In Fig.~\ref{4} the first result for the solution to the BKe is shown. There 
it can be seen how the amplitude $N$ starts at zero for small transverse 
distances $r$ (color transparency) 
to reach the saturation regime $N \sim 1$ at $r$ about 4 $\rm GeV^{-1}$. 
This trend is general for all values of the rapidity veto but, as a new 
feature, we observe how the arrival of the saturation of the amplitude is 
delayed as the veto increases. This first plot was done for a rapidity of 
10 and a fixed coupling of $\alpha_s = 0.2$. 
In Fig.~\ref{5} we highlight how in a region of larger rapidity,  
${\rm Y} = 14$ saturation arises earlier in $r$ for the set of initial 
conditions we have chosen. In this case of larger center--of--mass energy 
the effect of the veto is more dramatic, considerably delaying the onset 
of saturation. 

\begin{figure}[htbp]
 \epsfig{file=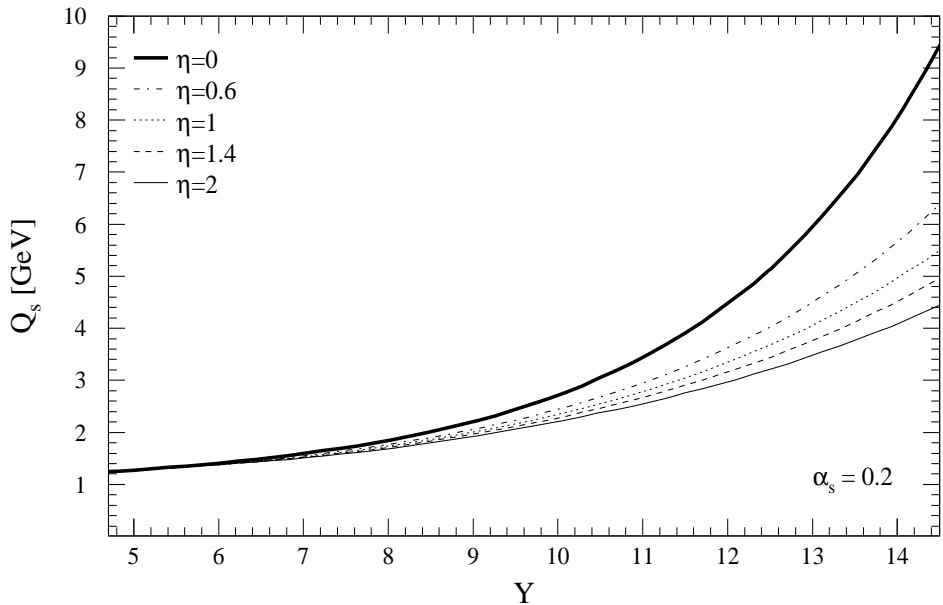,width=140mm, height=100mm}
\vspace{-0.5cm}
 \caption{Rapidity dependence of the saturation scale for fixed $\alpha_s$ 
and different vetoes.}
\label{7}
\end{figure}

\begin{figure}[htbp]
 \epsfig{file=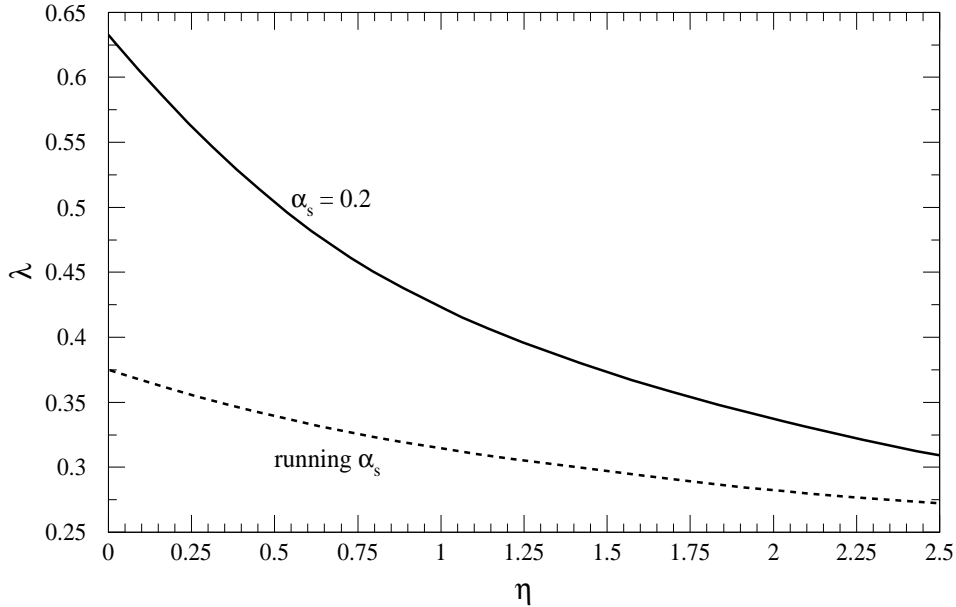,width=140mm, height=100mm}
\vspace{-0.5cm}
 \caption[]{$\eta$ dependence of $\lambda$  with running and fixed $\alpha_s=0.2$.}
\label{8}
\end{figure}

The major effect of the rapidity constraint comes when studying how the 
BK amplitude evolves with energy. As the intercept in the linear part 
is significantly reduced when higher order corrections are taken into 
account the saturation of the amplitude comes also later in rapidities. 
To illustrate this point we plot Fig.~\ref{6}, where we have chosen a 
typical value of $r = 0.75\,\,\rm GeV^{-1}$. 
Once again our estimated NLO corrections 
do delay the onset of saturation. To make this statement more quantitative 
we now proceed to study the saturation scale and its dependence with 
energy. Different definitions of the saturation scale can be associated 
with the step like function $N$~\cite{L}. They might lead to different 
normalizations although the energy dependence is qualitatively unique. 
For the sake of simplicity here we adopt the choice proposed in 
Ref.~\cite{LGLM} where it was taken at a point where $N$ reaches half:
\begin{figure}[htbp]
 \epsfig{file=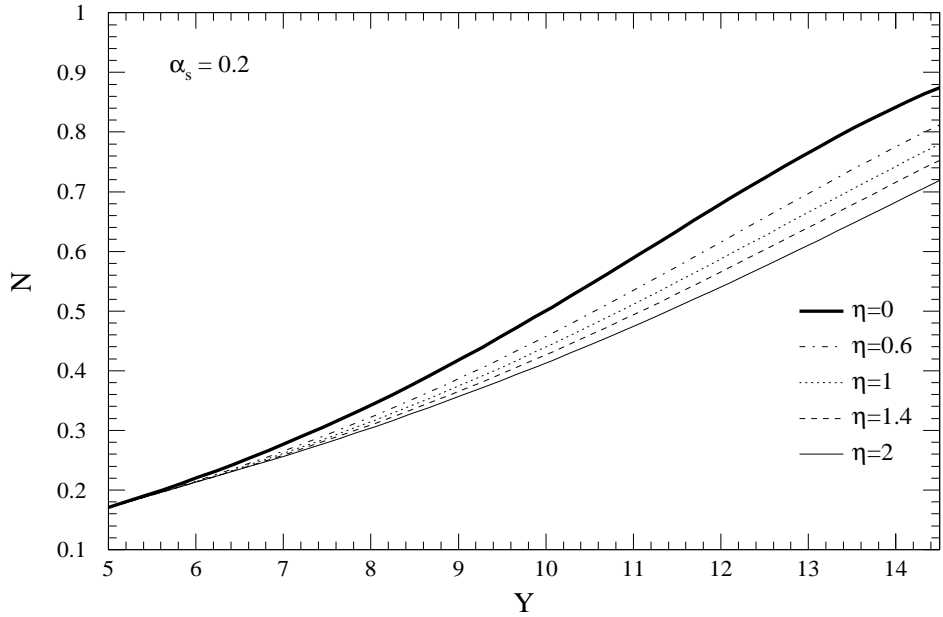,width=140mm, height=100mm}
\vspace{-0.5cm}
 \caption{Solution to BKe for different values of the veto 
as a function of the rapidity ${\rm Y}$, for $r=0.75\,\,\rm GeV^{-1}$ 
and $\alpha_s=0.2$ }
\label{6}
\end{figure}
\begin{eqnarray}
N(r_s,{\rm Y}) &=& \frac{1}{2}, 
\end{eqnarray}
with $r_s \equiv 2/Q_s$. For phenomenological applications the behavior of 
$\ln Q_s^2$ with rapidity ${\rm Y}$ can be fitted linearly  as in 
Eq.~(\ref{Ls}), i.e.,
\begin{eqnarray}
Q_s^2 &=& Q_0^2 \, e^{\lambda\,{\rm Y}}.
\end{eqnarray}
The numerical analysis of the rapidity dependence of this saturation 
scale is carried out in Fig.~\ref{7}. This plot reflects very clearly how 
saturation tends to appear later in rapidity, in particular, for the 
veto which reproduces the NLO intercept, $\eta \sim 2$. We have also 
performed a linear fit to estimate the linear power of Eq.~(\ref{Ls}), 
$\lambda$, this fit was done for phenomenological rapidities between 
${\rm Y} = 10 $
and  ${\rm Y} = 15$ so the expected value of the growth cannot be 
as large as in Fig.~\ref{Lambda_veto} due to preasymptotic 
effects~\footnote{The preasymptotic effects in saturation scale have been 
studied in Refs. \cite{MT,IIM,PM}. The 
numerical size of these terms is large at lower rapidities with the 
asymptotic values only reached at very large rapidities of the order of 
${\rm Y} \sim 100$ 
(see Ref. \cite{AW} for a similar  discussion).}. In fact, $\lambda$
is much smaller at the beginning of the evolution for rapidities up 
to $\rm Y \simeq 10$. 
In Fig.~\ref{8} we observe the transition power being of the order of 0.65 
already at zero $\eta$. This value is smaller than the equivalent obtained 
in Sec.~\ref{RapidityVeto}. Remarkably, the dependence on the rapidity veto 
is of the same functional form as in Fig.~\ref{Lambda_veto} with 
$\lambda$ reaching $\sim 0.31$ at a veto of $\eta = 2.5$.

The main conclusion of this section is to confirm the delay in energy space 
of the arrival of saturation when estimated higher order corrections are 
introduced in the BKe with fixed coupling constant. The numerical results 
when we introduce the veto on the BKe are consistent with those obtained 
from a more analytical approach in Sec.~\ref{RapidityVeto} based on the 
BFKL equation with a constraint in rapidity. In the following section 
the effect of this veto will be studied for phenomenological rapidities 
and running the QCD coupling.

\subsection{The running coupling case}

The BKe was originally derived for constant $\alpha_s$, the introduction 
of the running is part of the NLO corrections. At present the use of the 
running $\alpha_s$ in the BKe can only be done by modeling. In 
previous analysis introducing running seems to be phenomenologically 
favored by the data~\cite{GLLM}. This is because the effect of running 
$\alpha_s$ is to bring $\lambda$ down to about 0.3 in the 
phenomenologically relevant region of $x \ge 10^{-7}$. The main concern 
of this section is to study the stability of this value of 
$\lambda \sim 0.3$ when the veto is imposed on top of the running. 
 
Similar to the initial conditions,
$\alpha_s$ is taken at the leading order running with respect
to the external scale $4/r^2$. At large distances $\alpha_s$ is frozen
at the value $\simeq 0.5$. We have checked that our results are not sensitive
to variations of this value.  

\begin{figure}[htbp]
 \epsfig{file=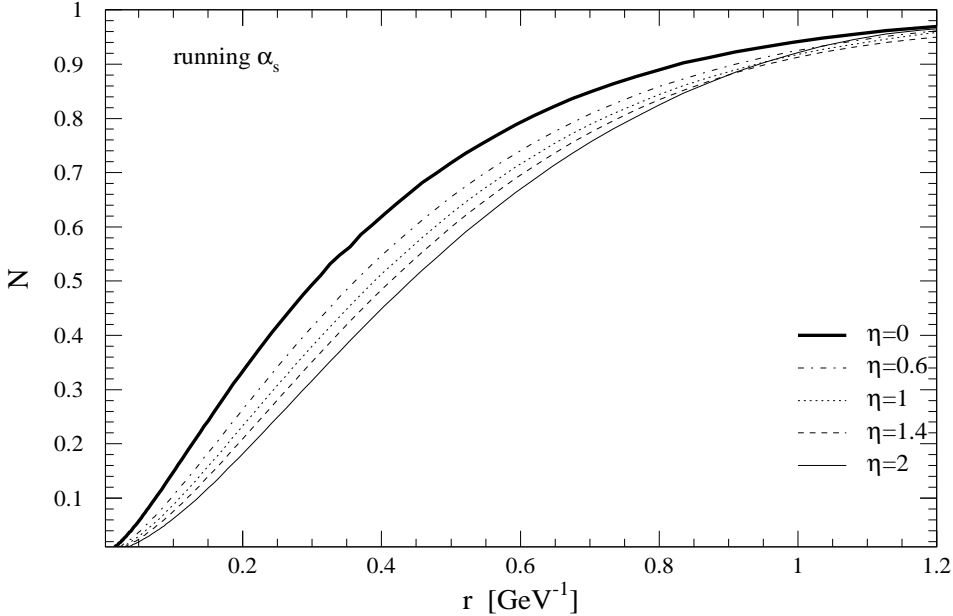,width=140mm, height=100mm}
\vspace{-0.5cm}
 \caption{Solution of the BKe for different vetoes as a function of
   $r$, for ${\rm Y}=14$ and running $\alpha_s$.}
\label{9}
\end{figure}

Let us start with Fig.~\ref{9} where we again show the region of small 
$r$ for the amplitude as calculated from the BKe introducing the rapidity 
constraint, this time running the coupling. As previously found the effect 
of the higher order corrections is to delay saturation. The rapidity 
chosen for this plot is 14. It is worth pointing out that the effect of the 
veto is reduced if we compare Fig.~\ref{9} to Fig.~\ref{5}, we will go 
back to this point soon below.

\begin{figure}[htbp]
 \epsfig{file=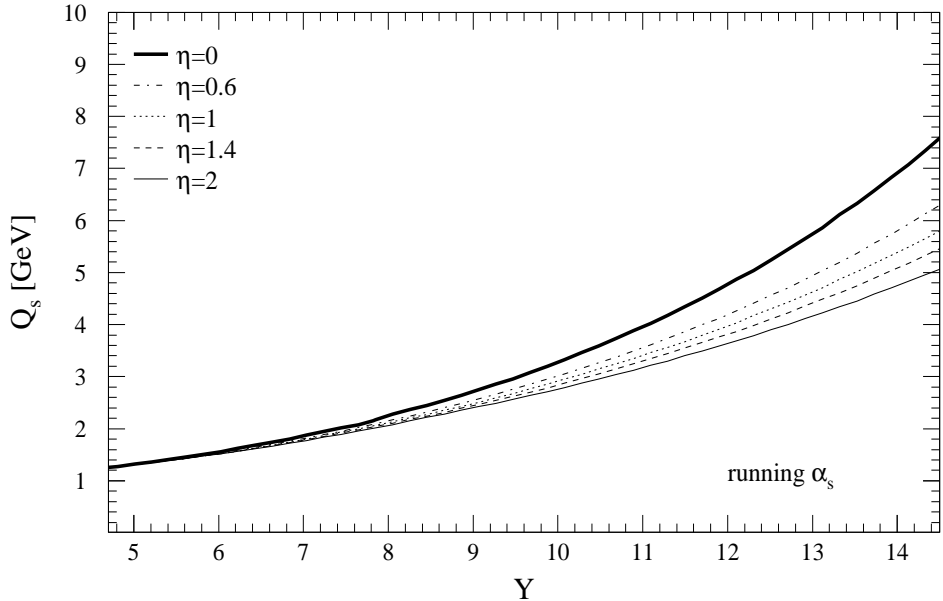,width=140mm, height=100mm}
\vspace{-0.5cm}
 \caption{{\rm Y} dependence of the saturation scale for running $\alpha_s$.}
\label{10}
\end{figure}

What about the energy dependence of the saturation scale? The answer to 
this question is plotted in Fig.~\ref{10} where the saturation scale is 
shown as a function of rapidity. The usual delay of the onset of saturation 
can be again observed although the effect of the estimated higher order 
corrections is smaller than for the case of fixed coupling in Fig.~\ref{7}. 
To make this more explicit we calculate the dependence of the $\lambda$ power 
in Eq.~(\ref{Ls}) as a function of the rapidity veto $\eta$ in the case 
of running coupling~\footnote{It is known from analytic studies that, 
contrary to the fixed coupling case, 
for running coupling  $\ln Q_s\sim \sqrt{Y}$. However, again due to 
large preasymptotic corrections~\cite{AW},  $\ln Q_s$ can be 
fitted linearly in rapidity for a limited range in ${\rm Y}$ relevant 
for phenomenology.} (Fig.~(\ref{8})).  The main 
conclusion is that the effect of the veto is not so big in the running 
coupling case, taking $\lambda$ from $\sim 0.37$ for $\eta = 0$ to 
about 0.27 for $\eta = 2.5$. This variation is much smaller than for the 
fixed coupling case.

The calculations in this section teach us that once the coupling is allowed 
to run the influence of other higher order corrections is diminished. The 
prediction for the growth of the saturation scale with energy remains of the 
order of $\lambda \sim 0.3$ for phenomenological energies independently 
of the  rapidity veto.

As a final remark, we have checked that the solutions to the BKe
for both  fixed and running $\alpha_s$ cases exhibit the geometrical
scaling property \cite{SGBK}. Namely, the amplitude $N(r,\rm Y)$ 
is a function of
the product $\tau\,=\,r\,Q_s(\rm Y)$, that is $N(\tau)$. Scaling holds in the 
saturation domain $\tau > 2$ and  extends to a much 
broader region  $\tau << 2$. For the BKe with $\eta=0$ 
scaling was shown in Refs. \cite{Braun2,L,GMS}. 
We have found that the scaling
is still present after introducing a non-zero veto, suggesting the full NLO
amplitude to be also a scaling function.   
   
\section{Conclusions}

In this work higher order corrections to the Balitsky--Kovchegov equation 
have been estimated. This estimate has been based on the introduction of 
a so--called ``rapidity veto'', which forbids two emissions to be very 
close in rapidity. It is known from Ref.~\cite{Schmidt:1999mz,Forshaw:1999xm} 
that the introduction of a veto, $\eta$, of $\sim 2.5$ units of rapidity 
mimics the Pomeron intercept predicted 
by other resummations of the NLO BFKL equation. We have estimated 
these higher order corrections first using analytical arguments imposing 
the rapidity veto on the LO BFKL equation obtaining a power growth of the 
saturation scale of $\lambda \simeq 0.45$, for $\alpha_s = 0.2$,
 consistent with that calculated in 
Ref.~\cite{Khoze:2004hx} and larger than that obtained in Ref.~\cite{DT}, 
our main result being Eq.~(\ref{critical_solution}), Eq.~(\ref{Ls}) and 
shown in Fig.~\ref{Lambda_veto}.

We have then pursued a numerical analysis of the introduction of the 
rapidity constraint in the full Balitsky--Kovchegov equation for 
phenomenological rapidities, without using asymptotic arguments. For a 
fixed coupling constant of 0.2 we observe that the power $\lambda$ decreases 
from $\sim 0.65$ for zero veto to $\sim 0.31$ for $\eta = 2.5$, with a 
dependence on the veto very similar to the previous analytical study, the 
main result plotted in Fig.~\ref{8}. 
When running coupling effects are also taken into account the effect 
of imposing the veto is not so important taking $\lambda$ at no veto 
from $\sim 0.37$ to $\sim 0.27$ for $\eta = 2.5$ (Fig.~\ref{8}).

As follows from the present analysis the running coupling effects account 
for the bulk of the NLO corrections to the BKe (see Fig.~\ref{8}). 
Given that, 
as shown in \cite{AW}, the dipole amplitude is not sensitive to the way 
the running is introduced, this suggests that phenomenological analysis 
including running coupling effects, as in \cite{GLLM}, do provide reliable 
predictions.

\hspace{-0.8cm} {\bf Acknowledgements:}

We would like to thank Nestor Armesto, Jochen Bartels, Jeff Forshaw, 
        Krzysztof Kutak, Eugene 
Levin, Lev Lipatov, Misha Ryskin, Anna Stasto and James Stirling for fruitful 
discussions and interest in this work. A.S.V. would like to thank the 
CERN Theory Division for hospitality, his contribution to this 
work was supported by an Alexander von Humboldt Postdoctoral
Fellowship. G.C. is supported by the \textit{Graduiertenkolleg}
``Zuk\"unftige Entwicklungen in der Teilchenphysik".


\begin{thebibliography}{0}

\bibitem{FKL}
L.~N.~Lipatov,
Sov.\ J.\ Nucl.\ Phys.\  {\bf 23} (1976) 338, 
E.~A.~Kuraev, L.~N.~Lipatov, V.~S.~Fadin,
Sov.\ Phys.\ JETP {\bf 45} (1977) 199, 
I.~I.~Balitsky, L.~N.~Lipatov,
Sov.\ J.\ Nucl.\ Phys.\  {\bf 28} (1978) 822.

\bibitem{MU97} A.~H.~Mueller,
{\it Phys. Lett.}  {\bf B 396} (1997) 251.

\bibitem{GLR} L. V. Gribov, E. M. Levin, and M. G. Ryskin, {\it Nucl. Phys.} 
{\bf B 188}  
(1981) 555; {\it Phys. Rep.} {\bf 100} (1983) 1.  
 
\bibitem{Mc}
L.~D.~McLerran,
{\it Lect.\ Notes Phys.}  {\bf 583} (2002) 291; hep-ph/0104285.

\bibitem{Golec-Biernat:1998js}
K.~Golec-Biernat and M.~Wusthoff,
Phys.\ Rev.\ D {\bf 59} (1999) 014017.

\bibitem{MUQI}  
A. H. Mueller and J. Qiu, {\it Nucl. Phys.} {\bf B 268} (1986) 427.  

 \bibitem{BA}  
Ia. Balitsky, {\it Nucl.Phys. } {\bf B 463}  (1996) 99.  

\bibitem{KO}  
Yu. Kovchegov, {\it Phys. Rev.} {\bf D 60} (2000) 034008.  

\bibitem{MU94}  
A. H.  Mueller, {\it  Nucl. Phys.}  {\bf B 415} (1994) 373.  

\bibitem{Braun} 
M. Braun, {\it Eur. Phys. J.} {\bf C 16} (2000) 337.  

\bibitem{BLV} 
J. Bartels, L.N. Lipatov, and J.P. Vacca, hep-ph/0404110. 

\bibitem{ILM}  
E. Iancu, A. Leonidov, and L. McLerran,  {\it Nucl.\ Phys.}\  {\bf A  692} (2001) 583.  
  
\bibitem{LT}  
Yu. Kovchegov,  { \it Phys. Rev.} {\bf D 61} (2000) 074018;  
 E. Levin and K. Tuchin, {\it Nucl. Phys.} {\bf B 573} (2000) 833;  
 {\it Nucl.\ Phys.} {\bf A 691} (2001) 779.  

\bibitem{KW} A. Kovner and U.A. Wiedemann,
{\it Phys. Rev.} {\bf D 66} (2002) 051502.

\bibitem{IIM} E.~Iancu, K.~Itakura and L.~McLerran,  
{\it Nucl.\ Phys.}  {\bf A  708} (2002) 327.  

\bibitem{PM} S. Munier and R. Peschanski, {\it Phys. Rev.} {\bf D 69}
(2004) 034008; {\it Phys. Rev. Lett.} {\bf 91} (2003) 232001; hep-ph/0401215.

\bibitem{GLLM}  E. Gotsman, E. Levin, M. Lublinsky, and U. Maor,  
{\it Eur. Phys. J} {\bf C 27} (2003) 411; parameterizations are available at
www.desy.de/$\sim$lublinm/.

\bibitem{Braun2} N. Armesto and M. Braun,  {\it Eur. Phys. J.} {\bf  
    C 20} (2001) 517.  

\bibitem{GMS} K. Golec-Biernat, L. Motyka, A. Stasto, {\it Phys. Rev.}  
  {\bf D 65} (2002) 074037.  

\bibitem{LGLM} M. Lublinsky, E. Gotsman, E. Levin, and U. Maor,  
{\it Nucl. Phys.} {\bf A 696} (2001) 851.  

\bibitem {Weigert} K. Rummukainen and H. Weigert, {\it Nucl. Phys.} {\bf A 739}
(2004) 183. 

\bibitem{BKEb} K. Golec-Biernat and  A.M. Stasto,
{\it Nucl. Phys.} {\bf B 668} (2003) 345;  \\
 E. Gotsman, M. Kozlov, E. Levin, U. Maor, and E. Naftali, 
hep-ph/0401021.

\bibitem{KS} K. Kutak and A. Stasto, hep-ph/0408117.

\bibitem{AW} J.~L.~Albacete, N.~Armesto, J.~G.~Milhano, C.~A.~Salgado and 
U.~A.~Wiedemann, hep-ph/0408216.

\bibitem{GLLMN}
E. Gotsman, E. Levin, M. Lublinsky, U. Maor, and E. Naftali,
{\it Acta Phys. Polon.} {\bf B 34} (2003) 3255; \\
J. Bartels, E. Gotsman, E. Levin, M. Lublinsky, and U. Maor,
{\it Phys. Rev.} {\bf D 68} (2003) 054008; {\it Phys. Lett.} 
{\bf B 556} (2003) 114; \\
E. Levin and M. Lublinsky, {\it Nucl. Phys.} {\bf A 696} (2001) 833;
{\it Phys. Lett.} {\bf B 521} (2001) 233.

\bibitem{L}
 M. Lublinsky,  {\it Eur. Phys. J.} {\bf C 21} (2001) 513.

\bibitem{Iancu} E. Iancu, K. Itakura and S. Munier {\it Phys.Lett.}
{\bf B 590} (2004) 199. 

\bibitem{Fadin:1998py}
V.~S.~Fadin, L.~N.~Lipatov,
Phys.\ Lett.\ B {\bf 429} (1998) 127.

\bibitem{Ciafaloni:1998gs}
M.~Ciafaloni, G.~Camici,
Phys.\ Lett.\ B {\bf 430} (1998) 349.

\bibitem{Fadin:2000nq}
V.~S.~Fadin,
Nucl.\ Phys.\ Proc.\ Suppl.\  {\bf 99A} (2001) 204.

\bibitem{Fadin:2002tu}
V.~S.~Fadin, D.~Y.~Ivanov and M.~I.~Kotsky,
Nucl.\ Phys.\ B {\bf 658} (2003) 156.

\bibitem{Fadin:1999de}
V.~S.~Fadin, R.~Fiore, M.~I.~Kotsky and A.~Papa,
Phys.\ Rev.\ D {\bf 61} (2000) 094005.

\bibitem{Bartels:2004bi}
J.~Bartels and A.~Kyrieleis, 
hep-ph/0407051, \\
J.~Bartels, D.~Colferai, S.~Gieseke and A.~Kyrieleis, 
Phys.\ Rev.\ D {\bf 66} (2002) 094017, \\
J.~Bartels, S.~Gieseke and A.~Kyrieleis, 
Phys.\ Rev.\ D {\bf 65} (2002) 014006, \\
J.~Bartels, S.~Gieseke and C.~F.~Qiao, 
Phys.\ Rev.\ D {\bf 63} (2001) 056014
[Erratum-ibid.\ D {\bf 65} (2002) 079902].

\bibitem{BB} I. Balitsky and A. Belitsky, {\it Nucl. Phys.} {\bf B 629}
(2002) 290.

\bibitem{DT} D.N. Triantafyllopoulos, {\it Nucl. Phys.} {\bf B 648} (2003)
293.

\bibitem{Khoze:2004hx}
V.~A.~Khoze, A.~D.~Martin, M.~G.~Ryskin and W.~J.~Stirling,
 hep-ph/0406135.

\bibitem{Schmidt:1999mz}
C.~R.~Schmidt,
Phys.\ Rev.\ D {\bf 60} (1999) 074003.

\bibitem{Forshaw:1999xm}
J.~R.~Forshaw, D.~A.~Ross and A.~Sabio Vera,
Phys.\ Lett.\ B {\bf 455} (1999) 273.

\bibitem{NLLpapers} 
G.P.~Salam, JHEP{\bf 8907} (1998) 19,\\ 
M.~Ciafaloni and D.~Colferai, Phys. Lett.{\bf B452} (1999) 372,\\ 
M.~Ciafaloni, D.~Colferai and G.P.~Salam, Phys. Rev. {\bf D60} (1999) 
114036,\\ 
R.S.~Thorne, Phys. Rev. {\bf D60} (1999) 054031,\\
J.~R.~Forshaw, D.~A.~Ross and A.~Sabio~Vera, 
Phys.\ Lett.\ B {\bf 498} (2001) 149,\\ 
G.~Altarelli, R.~D.~Ball and S.~Forte, Nucl.\ Phys.\ B {\bf 575} (2000) 313, Nucl.\ Phys.\ B {\bf 621} (2002) 359, Nucl.\ Phys.\ B {\bf 674} (2003) 459,\\ 
M.~Ciafaloni, D.~Colferai, G.~P.~Salam and A.~M.~Stasto, Phys.\ Lett.\ B {\bf 576} (2003) 143, Phys.\ Rev.\ D {\bf 68} (2003) 114003, Phys.\ Lett.\ B {\bf 587} (2004) 87,\\ 
J.~R.~Andersen and A.~Sabio Vera, Phys.\ Lett.\ B {\bf 567} (2003) 116, 
Nucl.\ Phys.\ B {\bf 679} (2004) 345, hep-th/0406009.

\bibitem{LL} E. Levin and M. Lublinsky, {\it Nucl. Phys.} {\bf A 730}
(2004) 191.

\bibitem{MT} A.H. Mueller and D.N. Triantafyllopoulos, 
{\it Nucl.\ Phys.}  {\bf B 640} (2002) 331.

\bibitem{SGBK} A.~M.~Stasto, K.~Golec-Biernat and J.~Kwiecinski,
{\it Phys.\ Rev.\ Lett.}  {\bf 86} (2001) 596.

\end{thebibliography}
\end{document}